%% file: main.tex
\documentclass[
 preprint,
superscriptaddress,
 amsmath,amssymb,
 aps,
 prr
]{revtex4-2}

\usepackage[colorinlistoftodos]{todonotes}
\usepackage{graphicx}
\usepackage{xcolor}
\usepackage{dcolumn}
\usepackage{bm}
\usepackage{silence}
\usepackage{relsize}
\usepackage{placeins}
\usepackage{hyperref}

\newcommand{\eye}{\mbox{$\mbox{1}\!\mbox{l}\;$}}

\usepackage{academicons}
\definecolor{orcidlogocol}{HTML}{A6CE39}

\usepackage{tikz,xcolor,hyperref}

\definecolor{lime}{HTML}{A6CE39}
\DeclareRobustCommand{\orcidicon}{%
	\begin{tikzpicture}
	\draw[lime, fill=lime] (0,0) 
	circle [radius=0.16] 
	node[white] {{\fontfamily{qag}\selectfont \tiny ID}};
	\draw[white, fill=white] (-0.0625,0.095) 
	circle [radius=0.007];
	\end{tikzpicture}
	\hspace{-2mm}
}

\foreach \x in {A, ..., Z}{%
	\expandafter\xdef\csname orcid\x\endcsname{\noexpand\href{https://orcid.org/\csname orcidauthor\x\endcsname}{\noexpand\orcidicon}}
}
\begin{document}

\title{Local versus global stability in dynamical systems with consecutive Hopf-Bifurcations}

\newcommand{\orcidauthorP}{0000-0002-3240-0442} 
\newcommand{\orcidauthorB}{0000-0003-1607-9748} 
\newcommand{\orcidauthorD}{0000-0002-3623-5341} 
\newcommand{\orcidauthorS}{0000-0003-2214-9271} 
\newcommand{\orcidauthorC}{0000-0003-4733-5257} 

\author{Philipp C. B\"ottcher\orcidP{}}
\email[Electronic mail: ]{p.boettcher@fz-juelich.de}
\affiliation{Forschungszentrum J\"ulich, Institute of Energy and Climate Research,
Systems Analysis and Technology Evaluation (IEK-STE), 52428 J\"ulich, Germany}  
\affiliation{Forschungszentrum J\"ulich, Institute of Energy and Climate Research,
Energy Systems Engineering (IEK-10), 52428 J\"ulich, Germany}

\author{Benjamin Schäfer\orcidB{}}
\affiliation{Karlsruhe Institute of Technology (KIT), Institute for Automation and Applied Informatics (IAI), Hermann-von-Helmholtz-Platz 1, 76344 Eggenstein-Leopoldshafen}

\author{Stefan Kettemann\orcidS{}}
\affiliation{Constructor University, Department of Physics \& Earth Sciences, Campus Ring 1, 28759 Bremen}
\affiliation{
Division of Advanced Materials Science, Pohang University of Science and Technology (POSTECH),San 31, Hyoja-dong, Nam-gu, Pohang 790-784, South Korea}

\author{Carsten Agert\orcidC{}}
\affiliation{DLR-Institute of Networked Energy Systems, Carl-von-Ossietsky Stra{\ss}e 15, 26129 Oldenburg, Germany}

\author{Dirk Witthaut\orcidD{}}
\affiliation{Forschungszentrum J\"ulich, Institute of Energy and Climate Research,
Systems Analysis and Technology Evaluation (IEK-STE), 52428 J\"ulich, Germany}  
\affiliation{Forschungszentrum J\"ulich, Institute of Energy and Climate Research,
Energy Systems Engineering (IEK-10), 52428 J\"ulich, Germany}
\affiliation{University of Cologne, Institute for Theoretical Physics, Z\"ulpicher Str. 77, 50937 Cologne, Germany}

\date{\today}

\begin{abstract}
Quantifying the stability of an equilibrium is central in the theory of dynamical systems as well as in engineering and control.
A comprehensive picture must include the response to both small and large perturbations, leading to the concepts of local (linear) and global stability.
Here, we show how systems displaying Hopf bifurcations show contrarian results on these two aspects of stability: Global stability is large close to the point where the system loses its local stability altogether. 
We demonstrate this effect for an elementary model system, an anharmonic oscillator and a realistic model of power system dynamics with delayed control. Detailed investigations of the bifurcation explain the seeming paradox in terms of the location of the attractors relative to the equilibrium.
\end{abstract}

\maketitle

\section{\label{sec:intro}Introduction}

Stability is an essential concept in the study of dynamical systems across disciplines \cite{strogatz2018nonlinear}. Given a perturbation, does a system relax back to a desired equilibrium state or not?
A loss of stability can have catastrophic consequences, as for instance the collapse of an ecosystem \cite{gross2009generalized}, the tipping of an element of the climate system \cite{lenton2008tipping} or a blackout of technical infrastructures such as the power grid \cite{witthaut2022collective}.  
Large perturbations are particularly hard to grasp, and one typically has to resort to extensive numerical simulations to assess the stability of an equilibrium. In this article, we demonstrate a surprising aspect of stability to large perturbations:
Certain systems are most stable when one expects the opposite, just before they become entirely unstable.

Traditionally, local stability has been central in the study of dynamical systems in the physical sciences. 
For a system in equilibrium affected by a small perturbation, the equations of motion can be linearized around its equilibrium point \cite{strogatz2018nonlinear}. The resulting Jacobian matrix gives a comprehensive picture of the dynamics in the neighborhood of the equilibrium according to the Hartman--Grobman theorem\cite{hartman1960lemma}. 
If all eigenvalues of the Jacobian matrix have a negative real part, then small perturbations will relax exponentially fast back to the equlibrium point. Thus we denote this equilibrium as linearly stable. 

Large perturbations are much more challenging to address as linearization around an equilibrium is no longer justified. 
In some cases it is possible to prove global stability in systems as diverse as neural networks and power systems \cite{arik2002analysis, wang2005global, selivanov2015adaptive,barabanov2017conditions}, but in many cases one has to resort to numerical investigations.
An important domain-independent concept to quantify the global stability is the basin of attraction\cite{ott2006basin} $\mathcal{B}$ -- the set of initial points in state space from which the system converges to a given attractor. 
The geometry of such a basin can be extremely complex, especially in large dimensions \cite{nusse1996basins}. 
Its volume, however, can be evaluated by numerical simulations: Drawing $E$ random initial conditions from a range of suitable initial conditions, the relative volume of the basin of attraction of one fixed point is estimated as $S_{\mathcal{B}} = M/E$, where $M$ is the number of initial conditions converging to that fixed point. 
If the sampling is extensive enough, the volume provides a quantitative measure of global stability, which can be interpreted as the likelihood to return to an attractor after a random perturbation \cite{wiley2006size,menck2013basin}. 
We will focus exclusively on the basin of fixed points and will not consider more complex attractors.

Local and global stability do not necessarily align \cite{michel2008stability}. 
Obviously, local (i.e., linear) stability is a necessary condition for a non-zero basin size -- but little can be said beyond this statement. 

In this article we demonstrate that local and global stability can even behave in a completely opposite way. 
We introduce a class of systems where the basin size assumes its maximum at a bifurcation point where linear stability is lost. This surprising behavior is demonstrated both for stylized models which allow for an analytic treatment and for advanced models inspired by engineering applications. All three systems share the same generic mechanism: The variation of an external control parameter induces a series of consecutive sub- and super-critical Hopf bifurcations. 

Overall, we have to understand that stability is a concept with multiple facets where linear stability and basin size may provide complementary information \cite{menck2013basin, schafer2016taming} and local stability implies global stability only under specific conditions \cite{chen2001global}. 
We note that extensions of network stability and synchronization often still rely on linear stability \cite{gambuzza2021stability}, while extensions of basin stability in terms of "survivability" are concerned with the transient behavior of the system towards a fixed point \cite{hellmann2016survivability} but do not provide further insights into the basin of attraction of individual fixed points. 
In the following, we focus on basin volume estimates to quantify the global stability of a system, and we use the terms local and linear stability as synonyms. 

The article is organized as follows: We first study a stylized model in Sec.~\ref{sec:proto} to introduce the basic mechanism. We then proceed to a more advanced model in Sec.~\ref{sec:kicked_system}, a kicked anharmonic oscillator, for which some analytic insights can be drawn by discretizing the dynamics. Finally, we analyze a dynamical system inspired by the load-frequency control in electric power engineering in Sec.~\ref{sec:real_life}. 
Taking into account delays in the control cycle, the system shows a similar series of Hopf bifurcations.

\section{Prototypical System}
\label{sec:proto}

\begin{figure}
    \centering
    \includegraphics[width=.475\columnwidth]{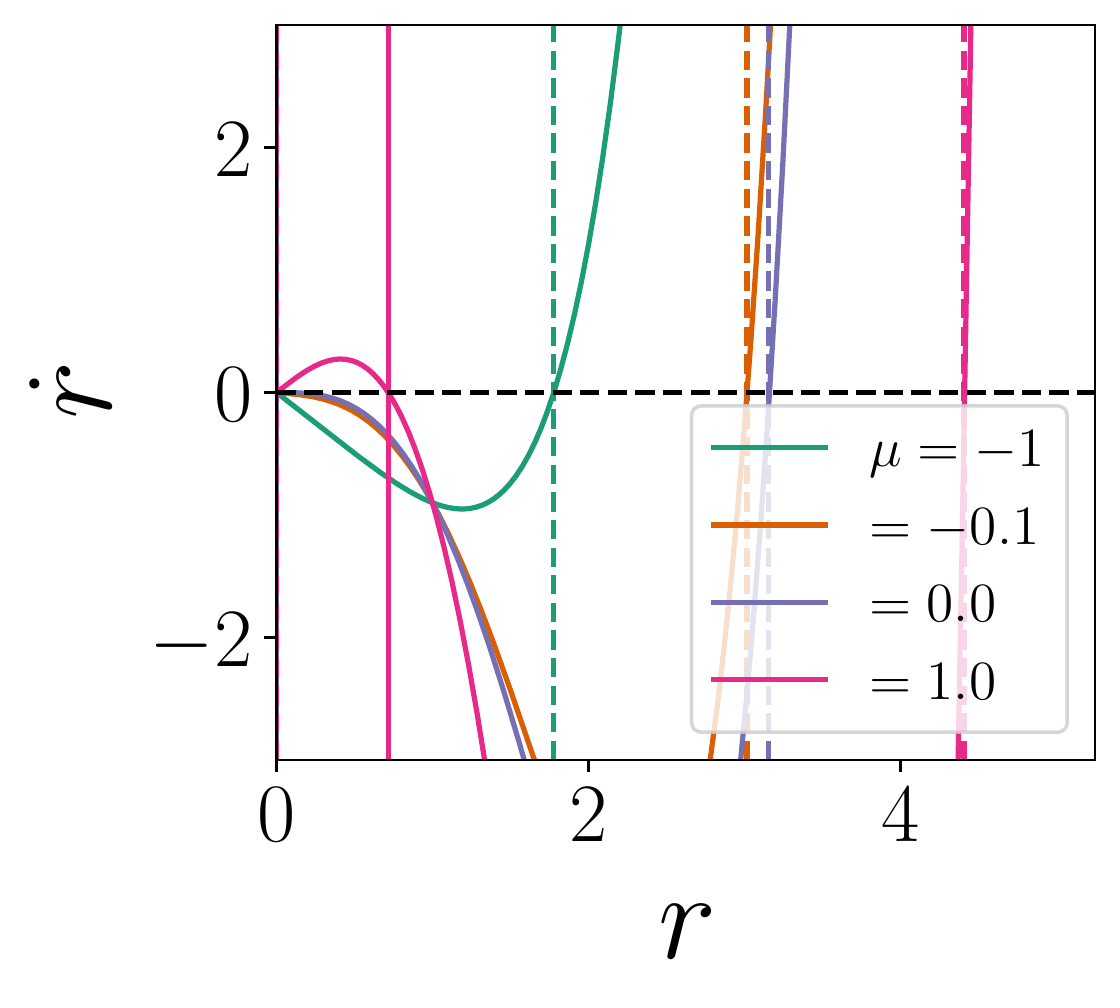}
    \includegraphics[width=.475\columnwidth]{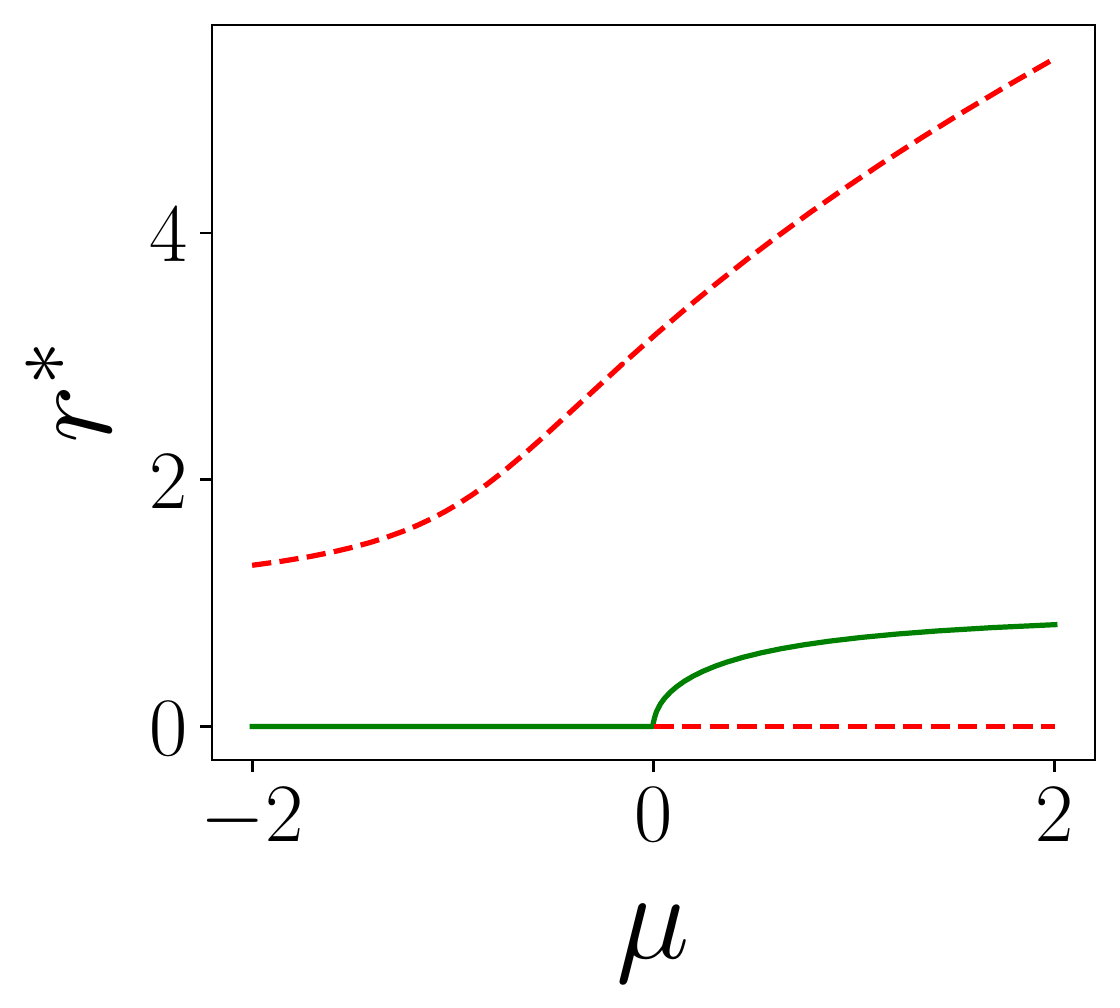}
    \caption{Attractors in the prototypical system \eqref{eq:proto}. 
    Left: Visualization of the equations of motion. Fixed points or limit cycles are found where $\dot r=0$ as indicated by the solid and dashed vertical lines for stable and unstable limit cycles, respectively.
    Right: Bifurcation diagram. The stable attractor is shown by the solid green line and the unstable attractors by the dashed red lines. A supercritical Hopf bifurcation occurs at $\mu=0$.
    }
    \label{fig:proto_rhs}
\end{figure}

We will show the basic mechanism of how local and global stability yield contradicting results using a stylized model which allows for a full analytic treatment. 
We consider a particle moving in the two-dimensional plane $\mathbb{R}^2$, generalizing the standard form of the Hopf bifurcation. 
Using polar coordinates with radius $r$ and angle $\psi$, the equations of motion read 

\begin{equation}
\begin{aligned}
    \dot{r} &= \mu r - (1+\mu) r^3 + \alpha r^5 \nonumber\\
    \dot{\psi} &= \omega + b r^2. 
    \label{eq:proto}
\end{aligned}
\end{equation}

Here, the dot denotes the differentiation with respect to time and $\mu$, $\omega>0$, $\alpha>0$ and $b>0$ are parameters. 
In the following, we analyze the system's dynamics when the parameter $\mu$ is varied while all other parameters are kept fixed. 
In all numerical examples we set $\alpha=0.1$ and $b=1$.

Attractors are found by setting $\dot r=0$.
The resulting attractors for the previously mentioned parameters are illustrated in Fig.~\ref{fig:proto_rhs}. 
The system always has a fixed point at $r^*=0$. 
Furthermore, limit cycles are found at the real positive roots of the polynomial equation
\begin{align}
    \mu - (1+\mu) r^2 + \alpha r^4 = 0.
\end{align}
Varying the value of the parameter $\mu$, we find the following scenario: For $\mu \le 0$, the fixed point $r^*=0$ is linearly stable and one unstable limit cycle exists at a radius
\begin{align*}
   r_2^2 =  \frac{1+\mu}{2\alpha} + 
           \sqrt{\frac{(1+\mu)^2}{4\alpha^2}-\frac{\mu}{\alpha}}.
\end{align*}
A supercritical Hopf bifurcation takes place at $\mu=0$. For $\mu>0$, the fixed point is unstable and a stable limit cycle exists at
\begin{align}
    r_1^2 = \frac{1+\mu}{2\alpha} -
           \sqrt{\frac{(1+\mu)^2}{4\alpha^2}- \frac{\mu}{\alpha}}.
\end{align}

\begin{figure}
    \centering
    \includegraphics[width=\columnwidth]{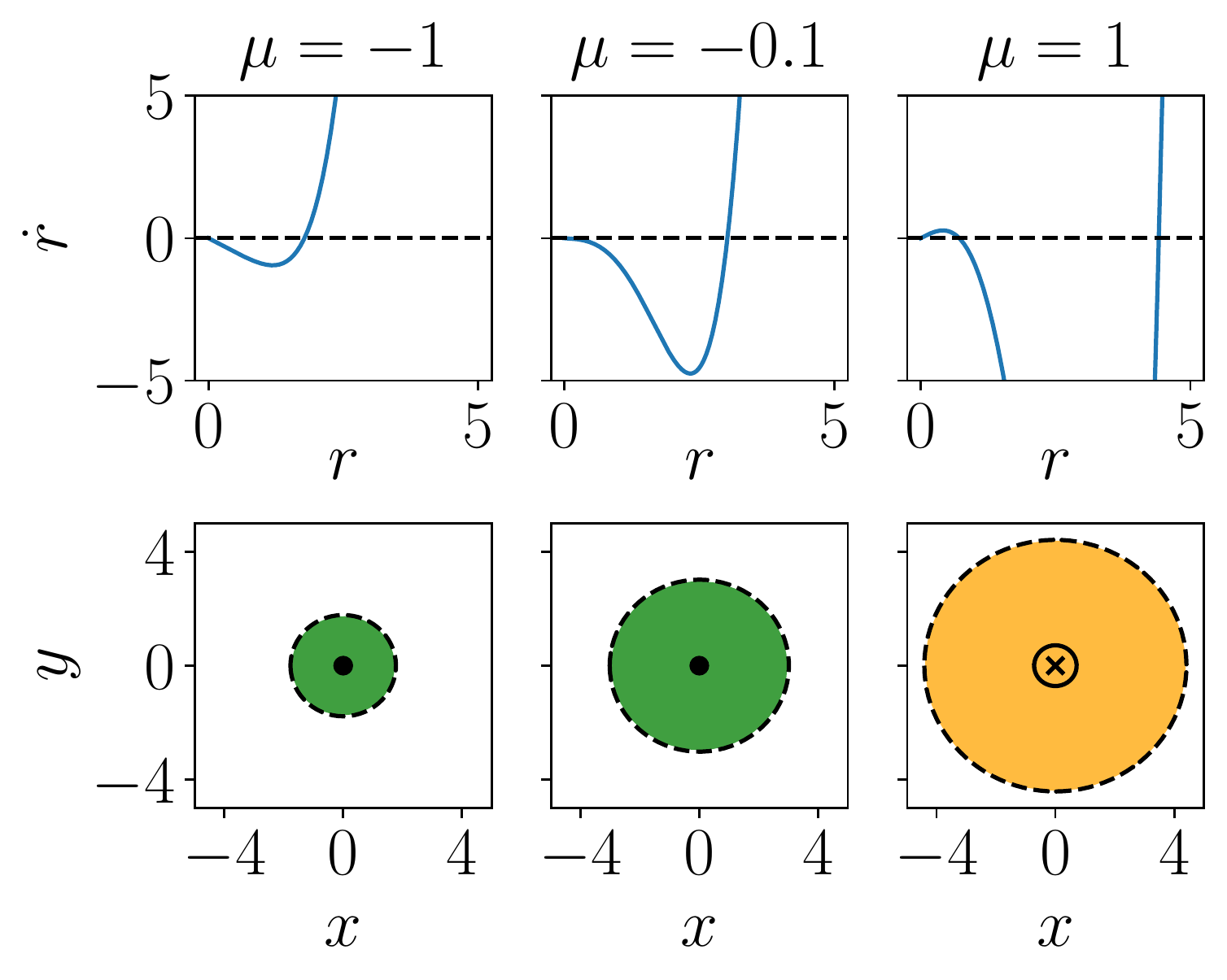}
    \caption{The basin of the fixed point is limited by the surrounding limit cycle.
    Upper panels: Visualization of the equations of motions \eqref{eq:proto} for $\mu=-1, -0.1, 1.$ (from left to right).
    Fixed points or limit cycles are found where $\dot r=0$.
    Lower panels: Fixed point (dot/cross) and limit cycle (solid/dashed lines) in the plane. 
    The green area shows the basin of attraction of the stable fixed point. 
    A Hopf bifurcation occurs at $\mu=0$, where the fixed point looses stability and a new stable limit cycle emerges. 
    The basin of the fixed point is bounded by the limit cycle, whose radius increases monotonically with the parameter $\mu$.
    For $\mu>0$ there is no basin of attraction since the fixed point is not stable.
    }
    \label{fig:proto_basin}
\end{figure}

Let us first consider the local stability of the fixed point $r^*=0$. Using Cartesian coordinates, the linearized dynamics reads
\begin{align}
    \frac{d}{dt}
    \begin{pmatrix}
       x \\ y
    \end{pmatrix}
    =
    \underbrace{
        \begin{pmatrix}
            \mu & - \omega \\ \omega & \mu
        \end{pmatrix}
    }_{=: \mathbf{J}}
    \begin{pmatrix}
       x \\ y
    \end{pmatrix} ,
\end{align}
and the eigenvalues of the Jacobian $\mathbf{J}$ are found as $\lambda_\pm = \mu \pm i \omega$. The real part, which encodes the linear stability of the fixed point, is directly given by the parameter $\mu$. The fixed point is stable for $\mu<0$ and stability is lost at $\mu=0$ in a Hopf bifurcation.

Now we turn back to the global stability. We can directly read of the basin of attraction of the fixed point $r^*=0$ for $\mu \le 0$,
\begin{align}
    \mathcal{B}_{r^*} = 
     \Big\{ \vec x \in \mathbb{R}^2 \;  \Big| \; 
    \| \vec x \| < r_2 \Big\}  
\end{align}
and we can use $r_2$ to quantify the basin size. For $\mu \ge 0$, the fixed point $r^*=0$ is unstable and we set the basin size to zero for the sake of convenience. 

The basin of attraction is shown in Fig.~\ref{fig:proto_basin} for three values of $\mu$ together with the limit cycles. We find that local and global stability behave in an opposite way. As $\mu$ increases from negative values towards zero, the unstable limit cycle moves outwards such that the basin of attraction of the fixed point grows.
However, the local stability of the fixed point weakens until it is lost in a Hopf bifurcation at $\mu=0$. 
Summarizing, the system reaches its maximal global stability (in terms of basin volume) at the same time when linear stability is lost at the bifurcation point.

\section{Driven Nonlinear Oscillator}
\label{sec:kicked_system} 

\begin{figure}[tb]
    \centering
    \includegraphics[width=\columnwidth]{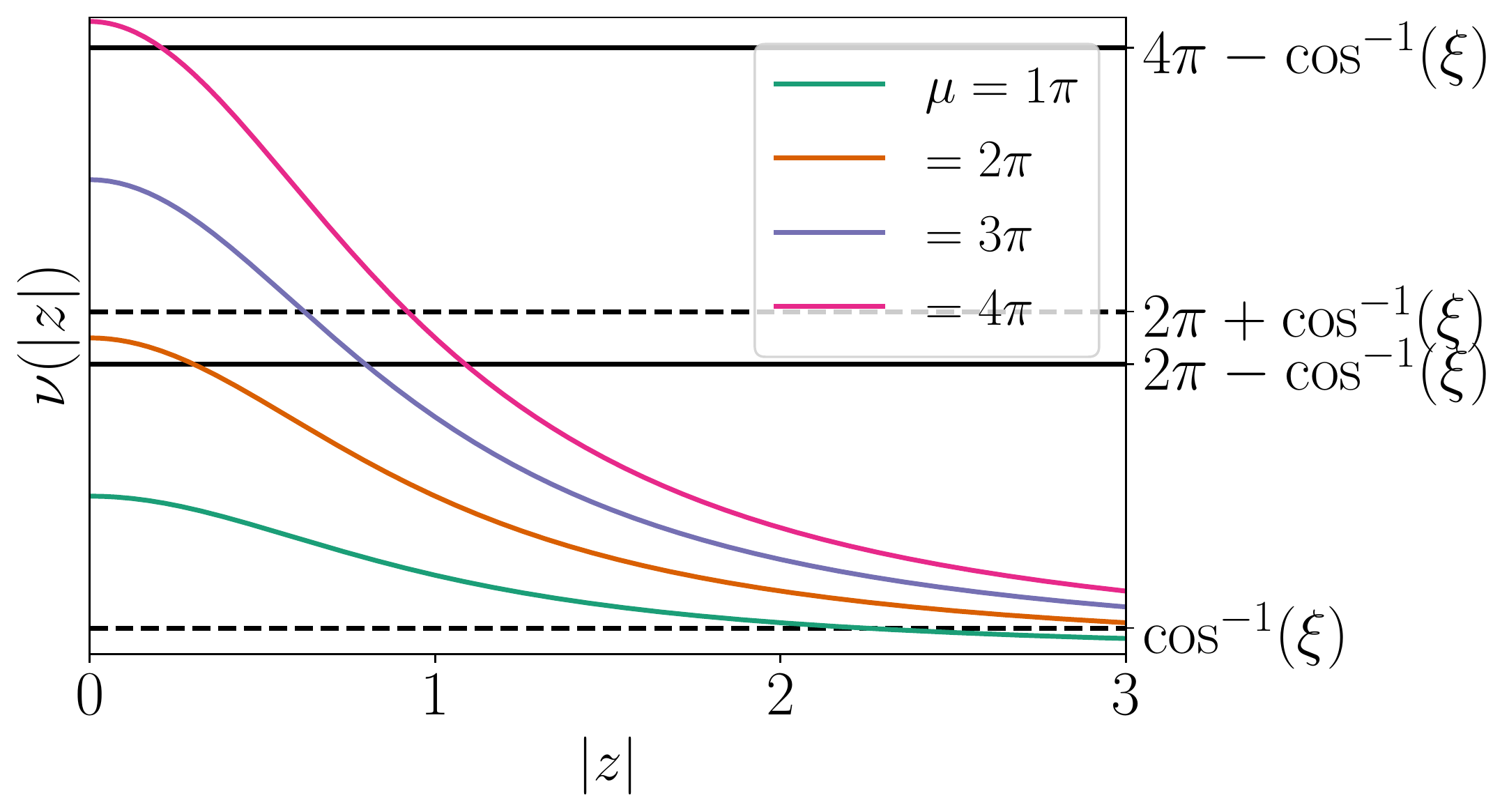}
    \caption{
    Understanding the bifurcations in the kicked anharmonic oscillator as a function of the control parameter $\mu$.
    Limit cycles are approximately determined by the solutions of Eq.~\eqref{eq:kicked_cosnu}. 
    Unstable limit cycles are found for $\nu(|z|) = 2 \pi n +\cos^{-1}(\xi)$ and stable limit cycles for $\nu(|z|) = 2 \pi n -\cos^{-1}(\xi)$ with the abbreviation $\xi=(1+\gamma^2-c^2)/{2\gamma}$ and $n \in \mathbb{N}$. 
    The figure shows the function $\nu(|z|)$ for different values of the control parameter $\mu$. 
    Stable (unstable) limit cycles exist where the function crosses the solid (dashed) horizontal lines.
    }
    \label{fig:kicked_scheme}
\end{figure}

The contrarian behavior of global and local stability can generally be found in systems where limit cycles and fixed points coexist.
This scenario can occur repeatedly in a system with consecutive sub- and supercritical Hopf bifurcations, as we will now demonstrate for a more general model system. In particular, we examine a damped and driven anharmonic oscillator. 
We assume that the complex amplitude $z \in \mathbb{C}$ evolves according to the equations of motion
\begin{align}
    \dot{z} &= (i \omega(|z|) - \hat{\gamma}) z + g(|z|,t),  \label{eq:kicked_eom}
\end{align}
with $\hat{\gamma}$ being the damping constant and $i$ the imaginary unit. We assume that the oscillator is anharmonic, such that the frequency $\omega$ decreases monotonically with the amplitude $|z|$. The driving function is periodic in time, i.e. $g(|z|,t+T) = g(|z|,t)$ for a given period $T\in \mathbb{R}$. Furthermore, the strength increases with the amplitude such that $g(0,t) = 0$. 

The dynamical system \eqref{eq:kicked_eom} always has a trivial fixed point $z^*=0$, which we interpret as the desired equilibrium state. The stability of this fixed point -- both local and global -- crucially depends on the existence of limit cycles. We will analyze this relation in detail for two special realizations of the external driving.

\subsection{Kicked system}

We now consider the case of a periodically kicked system, which allows for an approximate analytical solution. 
The nonlinear driving term reads
\begin{align}
    g(|z|,t) = a \sum\limits_{n=-\infty}^{+ \infty} \delta(t - n T) |z| 
    \label{eq:kick_function}
\end{align}
with an amplitude $a>0$. 
The nonlinear kicking is interpreted as
\begin{align}
    z(nT+\epsilon)-z(nT-\epsilon)
    = a |z(nT+\epsilon)|,
    \quad \epsilon \rightarrow 0.
\end{align}
Furthermore, we assume that the amplitude of the anharmonic oscillator decreases with the amplitude as
\begin{align}
    \omega(|z|) &= \frac{\mu}{T (1 + |z|^2) } \, .
    \label{eq:nu_amplitude}
\end{align}
We will analyze the resulting dynamics as a function of the control parameter $\mu$.

If the damping constant $\hat{\gamma}$ is sufficiently small, we can simplify the dynamics by assuming that the amplitude and thus the frequency $\omega$ remains approximately
constant between two kicks. Then we obtain
\begin{align}
    z((n+1)T-\epsilon) \approx
    \exp \{ (i \omega - \hat{\gamma}) T \} \,
    z(nT+\epsilon)
\end{align}
Defining  $z_n = z(nT+\epsilon)$, we thus obtain a discrete map 
\begin{align}
    z_{n+1} &= e^{i \nu(|z_n|)} \cdot \gamma \cdot z_n + c |z_n| \label{eq:kicked_map}
\end{align}
with $\nu(|z_n|) = \omega(|z_n|)T$, $\gamma = e^{-\hat{\gamma} T} \in [0,1]$ and $c = \gamma a>0$. 

Limit cycles with period $T$ are found by evaluating the condition $z_{n+1} = z_n$. Writing $z = |z| e^{i\alpha}$, the fixed point equation reads
\begin{align}
    e^{i\alpha} |z|  = \gamma  |z| e^{i (\alpha + \nu(|z|) )} + c |z|
\end{align}
For the non-trivial limit cycles we can solve this equation for the amplitude and phase and obtain
\begin{align}
    \cos( \nu(|z|) ) &= \frac{1+\gamma^2-c^2}{2\gamma} 
    \label{eq:kicked_cosnu}
    \\
    \cos( \alpha ) &= \frac{1-\gamma^2+c^2}{2c}.
\end{align}
Since real solutions only exist if the right-hand side of both equations is in the interval $[-1,+1]$, we assume this from now on. 
Note, the function $\nu(|z|)$ critically determines whether limit cycles exist or not. 
For the function given in Eq.~\eqref{eq:nu_amplitude}, we find the following behavior:
For $\mu=0$, we typically find no solution to Eq.~\eqref{eq:kicked_cosnu} and thus no limit cycle. As $\mu$ increases, additional solutions come into being as illustrated in Fig.~\ref{fig:kicked_scheme}. 
The emerging limit cycles are alternately unstable and stable and their amplitude $|z|$ increases monotonically with $\mu$.

As a consequence, the phase space of the kicked system shows a pronounced shell structure for large values of $\mu$. The trivial fixed point $z^*=0$ is surrounded by stable and unstable limit cycles. These limit cycles move outwards as $\mu$ increases and new cycles emerge repeatedly via Hopf bifurcations. The fixed point becomes unstable when a new stable limit cycle emerges and it becomes stable again when a new unstable limit cycle emerges. 

\begin{figure}[tb]
    \centering
    \includegraphics[width=\columnwidth]{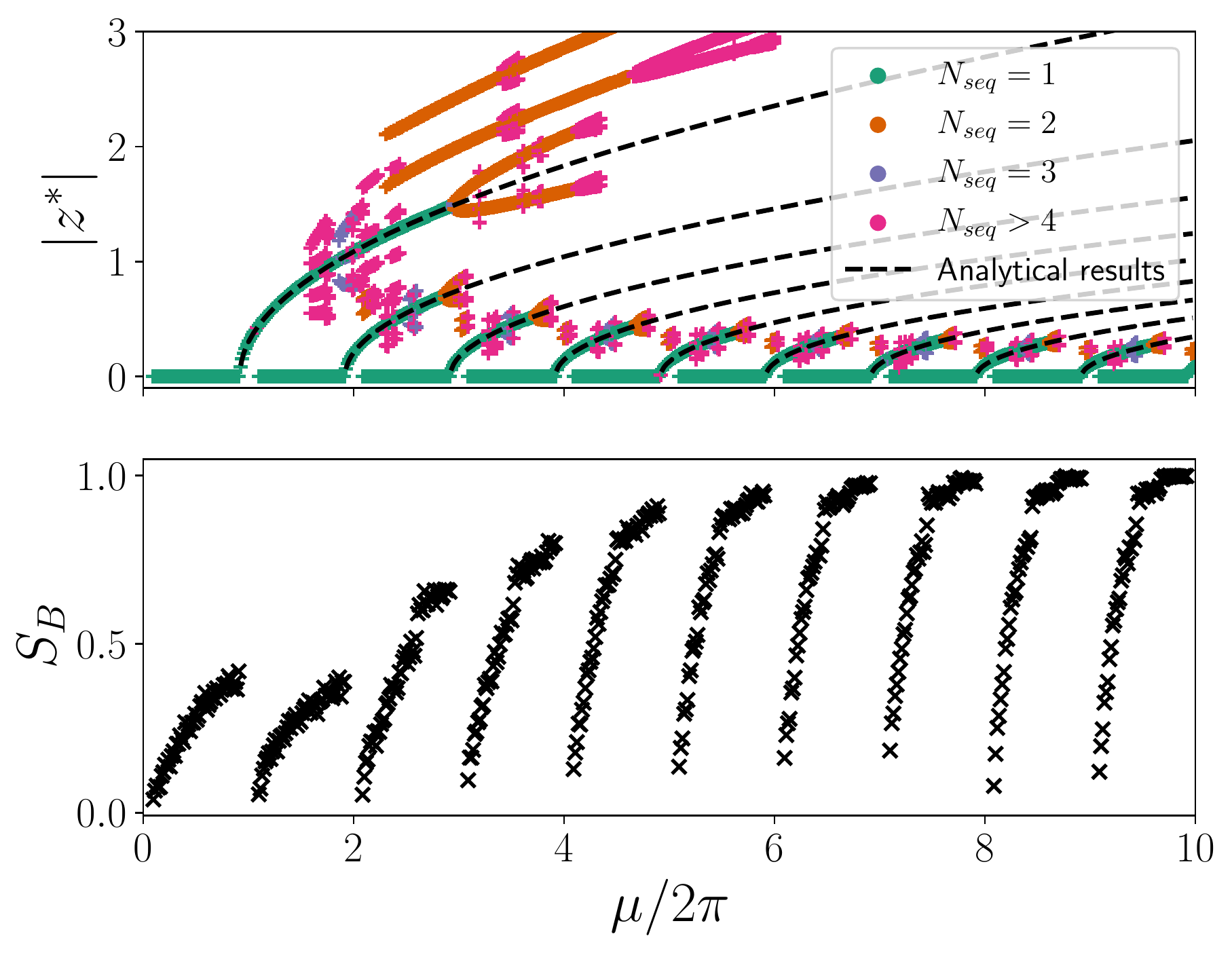}
    \caption{
    Attractors and basin stability $S_\mathcal{B}$ of the discrete map (which approximates the kicked anharmonic oscillator \eqref{eq:kicked_eom}) as a function of the control parameter $\mu$.
    Upper panel: Absolute value of the stable limit cycles with period $N_{seq}$. 
    The trivial fixed point $z^*=0$ undergoes a series of consecutive sub- and super-critical Hopf bifurcations, where it changes from stable to unstable and vice versa. 
    The positions of the non-trivial stable limit cycles, calculated by solving Eq.~\eqref{eq:kicked_cosnu}, are indicated by the dashed black lines.
    Lower panel: The basin stability $S_{\mathcal{B}}$ of the trivial fixed point increases monotonically with $\mu$ until it jumps to zero when stability is lost at a bifurcation point.}
    \label{fig:kicked_map_fxp_basin}
\end{figure}

The phase space structure determines the system's global stability. The basin of attraction of the fixed point is bounded by the nearest unstable limit cycle, whose size grows monotonically with $\mu$ -- until the next bifurcation takes place. 
Based on these consideration, we expect that local and global stability behave in an opposite way and that the basin size assumes its maximum right before local stability is lost.

We test these qualitative statements by numerical simulations. 
We scan the parameter $\mu$ in the range $[0,20\pi]$ and simulate the discrete dynamics given by Eq.~\ref{eq:kicked_map}. 
To check which states run into different attractors, we chose the random complex initial condition $z_0 = z_{0,r} + z_{0,i} i$ by uniformly sampling the real part $z_{0,r}$ and the imaginary part $z_{0,i}$ from $[-4, 4]$.
The map is iterated for $t_n=1000$ steps to check whether the discrete dynamics runs into an attractor.
Since we are interested in how the trivial fixed point at $z^*=0$ is affected by the other attractors, we quantify its global stability by counting the number of initial conditions that run into this fixed point, i.e. $S_{\mathcal{B}}$ is the fraction of initial conditions that returns to $z^*$.

To validate these consideration, we simulated the discrete map given by Eq.~\eqref{eq:kicked_eom}.
Setting the parameters to $c=0.5$, $\gamma=0.9$ and $T=1$, we find that the simulations confirm the expected behavior (see Fig.~\ref{fig:kicked_map_fxp_basin}). 
As $\mu$ increases, the trivial fixed point repeatedly switches from stable to unstable as new limit cycles emerge. During the stable intervals, the basin size $S_{\mathcal{B}}$ of the fixed point increases monotonically with $\mu$ and assumes its maximum at the bifurcation point. 
Notably, the limit cycles undergo further bifurcations which are not treated here as we focus on the stability of the fixed point.

\subsection{Continuously driven system}

\begin{figure}[tb]
    \centering
    \includegraphics[width=\columnwidth]{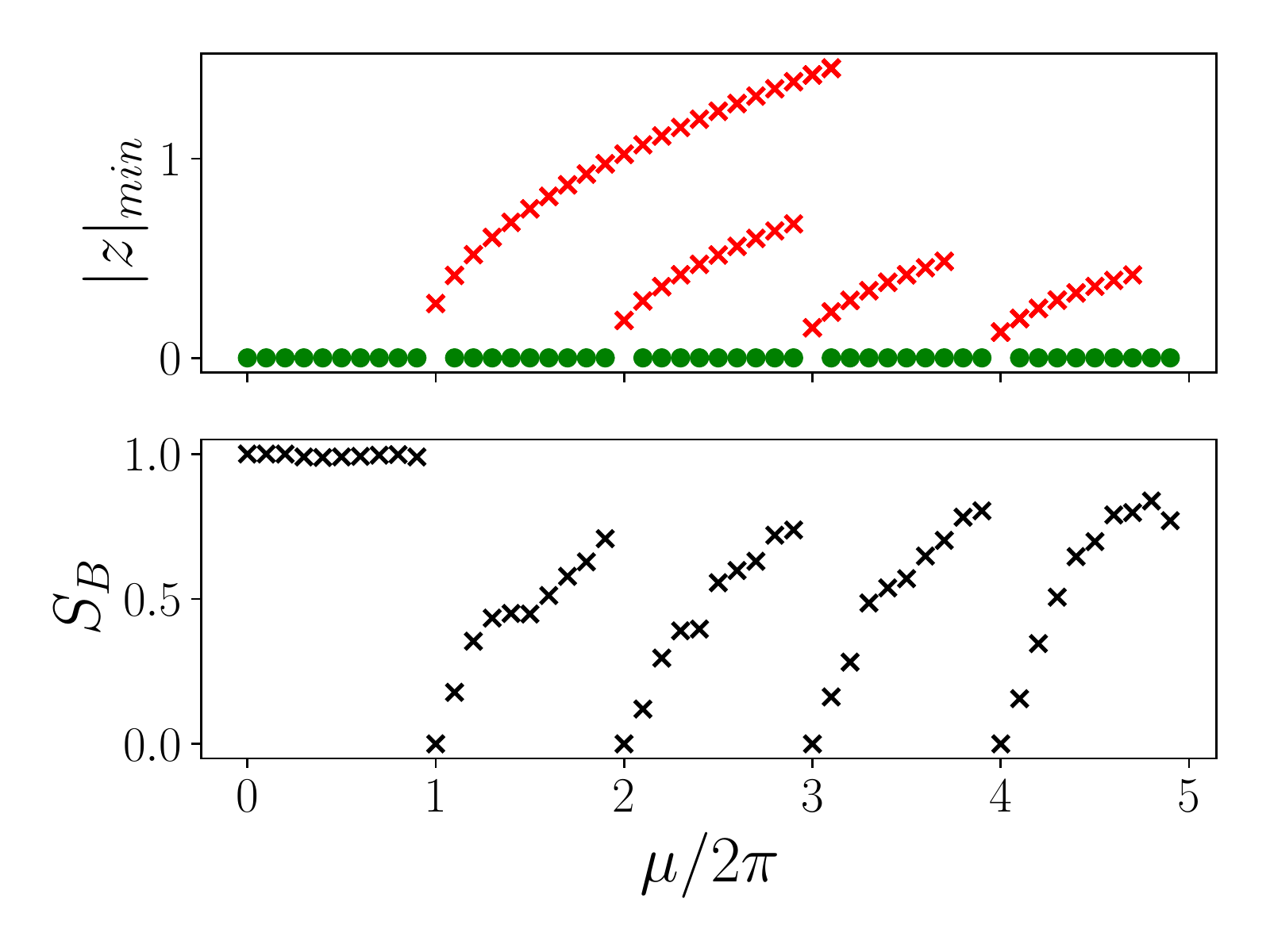}
    \caption{
    Fixed points and limit cycles of the driven anharmonic oscillator \eqref{eq:kicked_eom} as a function of the control parameter $\mu$. 
    Top panel: For each stable fixed point or limit cycle, we plot the minimum of the magnitude $|z|_{min}$. 
    Green dots indicate where the trivial fixed point is stable and red crosses show where stable limit cycles with a period of $T$ were detected. 
    Bottom panel: Basin stability $S_{\mathcal{B}}$ for the trivial fixed point. 
    The periodic driving is given by Eq.~\eqref{eq:driving_cont}. 
    }
    \label{fig:kicked_continuous}
\end{figure}

To show that the analytical and numerical results also hold for a time continuous system, we return to the original continuous dynamical system given in Eq.~\eqref{eq:kicked_eom} and assume a continuous driving.
We replace the delta function or more specifically the \emph{kick} by using a Gauss function $\Lambda(t) = (2\pi)^{-\frac{1}{2}} \cdot \exp{(-\frac{t^2}{2})}$, which results in
\begin{align} 
    g(|z|,t) &= \sum\limits_{n=0}^{\infty} N_{d} \cdot \Lambda \left(N_d \cdot \left[t - 
n T\right]\right) \cdot |z| 
    \label{eq:driving_cont}
\end{align}
where $N_d \in \mathcal{N}$ is a positive constant. 
Using this substitute, we study the dynamics numerically as a function of the control parameter $\mu$.
We sample uniformly in $[-3, 3]$ to get both the real and imaginary part of the inital conditions $z_0$. 
Choosing the same parameters, i.e. $c=0.5$, $\gamma=0.9$ and $T=1$, as in the discrete map, we solve the ODE in the time interval $t\in [0, 200]$.
Only a short range (i.e. $t_c = 20$) at the end of the resulting time series is used to evaluate whether or not the trivial fixed point is reached or if the dynamics end in a limit cycle.
The size of the basin $S_{\mathcal{B}}$ is again given as the fraction of initial state that end in the trivial fixed point.
The simulation results (see Fig.~\ref{fig:kicked_continuous}) confirm the qualitative picture obtained for the discrete map. 
In particular, we again find the opposite behavior of the local and global stability of the trivial fixed point.
Again the size of the basin of attraction increases with the diameter of the limit cycle and is largest slightly before the fixed point loses stability and a new limit cycle with small diameter appears.
Note, as in the case of the discrete map, there are more bifurcations present in the system that are hard to detect numerically. 
The limit cycles presented in Fig.~\ref{fig:kicked_continuous} are only the ones that have a the period of $T=1$ and follow the the same path in phase space.

\section{Oscillators with delayed control}
\label{sec:real_life} 

Hopf bifurcations often occur in delayed dynamical systems. Such a delay can arise in a feedback or control loop, where measurements and information processing requires some time. We will now demonstrate that the previously introduced scenario of consecutive Hopf bifurcations may occur in a control system of immense practical importance: the load-frequency control of electric power systems\cite{handbook2004policy,handbook2004appendix,  handbook2009policy}. 

\subsection{Power systems dynamics and control}

We consider an aggregated model of an electric power grid\cite{machowskipower, ulbig2014impact}, decomposing the grid into certain regions or control areas. The areas are strongly coupled internally, such that local  differences of the grid frequency are negligible. Hence, every area $i = 1,\ldots,N$ is characterized by its voltage phase angle $\theta_i(t)$ and the frequency deviation $\omega_i(t) = \dot \theta_i$ from the reference frequency $\omega_0$.
Ideally, all areas should run at the same nominal reference frequency $\omega_0 = 2\pi \times 50$ Hz or $\omega_0 = 2\pi \times 60$ Hz, but imbalances of power generation and load induce deviations. The load-frequency control measures these frequency deviations and adapts the generation to restore the balance and limit deviations from the reference state.

The dynamics of the aggregated model is described by the aggregated swing equation 
\begin{align}
   \dot \theta_i &= \omega_i \\
   A_i \, \dot \omega_i + k_{l,i} \, \omega_i(t) 
   &= P_{0,i}(t) + P_{c,i}(t) - \sum_{j} P_{ij}(t) , 
   \label{eq:2ndorder} 
\end{align}
using a frame of reference rotating at the frequency $\omega_0$. Here, $A_i$ quantifies the amount of inertia and $k_{l,i}$ is a damping constant due to generator damper windings or frequency-dependent loads. 
Effectively, each area is modeled as an aggregated synchronous machine\cite{ulbig2014impact}.
The right-hand side includes the balance of scheduled generation and load $P_{0,i}(t)$, the contribution of the load-frequency control system $P_{c,i}(t)$ and the flow to other areas given by
\begin{align}
    P_{ij}(t) = C_{ij} \sin( \theta_i(t) - \theta_j(t) ).
    \label{eq:power_flows}
\end{align}
The control system continuously monitors the grid and adapts the power $P_{c,i}(t)$ to restore the desired grid operation. Here we focus on primary control, also referred to as frequency containment reserve (FCR), which is activated within seconds. 
Further control layers exist, which are activated on longer time scales and will be neglected here for the sake of clarity. 
Primary control can be described as a proportional controller, adjusting the power proportional to frequency deviations. Both measurement and communication as well as the activation of a reserve power plant require some time. We thus model primary control, following earlier work \cite{schafer2015decentral,schafer2016taming} as
\begin{align}
    P_{c, i}(t) = - k_{P,i} \, \omega_{i}(t - \tau). \label{eq:control}
\end{align}
In the following, systems with homogeneous gains $k_{p,i} = k_P$ and damping constants $k_{l, i} = k_l$ for all $i = 1, \ldots, N$  were considered. 

In summary, the dynamics is given by the delay differential equation (DDE)
\begin{align}
    \dot \theta_i &= \omega_i \\
    \begin{split}
   A \, \dot \omega_i + k_{l} \, \omega_i(t) 
   &= P_{0,i} - k_{P} \, \omega_{i}(t - \tau)\\
   &\quad- C_{ij} \sin( \theta_i(t) - \theta_j(t) ) , 
   \end{split}
   \label{eq:full_dde} 
\end{align}
where the control is delayed by $\tau$ and works against a detected frequency deviation $\omega$ proportionally to the gain $k_P$.  

\subsection{Fixed points and oscillations}

Ideally, the power grid should be in a fixed point where the power balances $P_{0,i}$ are fixed and all areas are perfectly synchronized,
\begin{align}
    \omega_i(t) &= \omega^* \\
    \theta_i(t) &= \theta^*_i + \omega^*  t.
\end{align}
Recall that the variables $\theta_i$ and $\omega_i$ are defined in a rotating frame with reference frequency $\omega_0$. The fixed point values are determined by the algebraic equations
\begin{align}
    P_{0,i} - (k_{l}  + k_{P}) \omega^* 
      = \sum_j C_{ij} \sin(\theta^*_i - \theta^*_j).
\end{align}
Summing over all areas $i= 1,\ldots,N$ we further obtain
\begin{align*}
    \omega^* = \frac{\sum_i P_{0,i} }{ \sum_i (k_{l}  + k_{P}) }.
\end{align*}

Disturbances of the power balance can cause transient deviations from this fixed point. A notable example of this behavior are inter-area oscillations \cite{Vanfretti2010,Klein1991, ucte2007final}, large scale oscillations of the phases $\theta_i(t)$ and the power flows \eqref{eq:power_flows} of the entire grid, potentially over thousands of kilometers. 
The eigenmodes are determined by the structure of the grid, with typical frequencies in the range of 0.1–10 Hz. 
Inter-area oscillations are typically damped out in minutes.

In exceptional contingency situations, oscillations may also grow leading to a loss of synchrony in the grid and eventually a blackout. For instance, such an instability may arise after the loss of several transmission elements in a cascade of failures, see \cite{ucte2007final} for an example.

\subsection{Linear stability analysis\label{sec:power_linstab}}

In this section, we discuss the linear stability of the  desired fixed point. 
To this end, we linearize the equations of motion \eqref{eq:2ndorder} as $\theta_i(t) = \theta^*_i + \omega^*  t + \alpha_i(t)$. 
To simplify the analysis we assume that the grid is balanced in total such that $\sum_i P_{0,i}=0$ and $\omega^*=0$. 

We then obtain the linearized equations
\begin{align}
    \dot \alpha_i &= \omega_i \\
    A_i \dot{\omega_i} &= - k_{l} \omega_i 
    - k_{P} \omega_{i,\tau} - 
    \sum_j L_{ij} \alpha_{j}. 
\end{align}
Here, we have used the shorthand $\omega_{i,\tau} = \omega_{i}(t-\tau)$ and dropped all time dependencies for the sake of brevity.
The coupling between the areas is described by the network Laplacian $\mathbf{L} \in \mathbb{R}^{N \times N}$ with elements 
\begin{equation}
    L_{ij} =\left\{ \begin{array}{l l l}
    - C_{ij} \cos(\theta_{i}^* - \theta_{j}^*) 
    & \; \text{if} \; & i \neq j\\
    \sum_{m \neq i} C_{im} \cos(\theta_{i}^* - \theta_{m}^*) 
    &  & i = j.
    \end{array} \right.
\end{equation}

For further analysis we define a state vector
\begin{align*}
    \vec x = (\alpha_1,\ldots,\alpha_N,
              \omega_1,\ldots,\omega_N)^\top
\end{align*}
and rewrite the linearized equations in a matrix form \begin{align}
    \mathbf{A} \dot{ \vec{x}}(t) = \mathbf{N} \vec{x}(t) + \mathbf{D} \vec{x}(t-\tau).
    \label{eq:eom-lin-x}
\end{align}
with the block matrices
\begin{align*}
    \mathbf{A} = \begin{pmatrix}
    \eye & 0 \\ 0 & \mathbf{\hat{A}}  
    \end{pmatrix}, \;
    \mathbf{N} = \begin{pmatrix}
    0 & \eye \\ -\mathbf{L} &  -\mathbf{K}_l
    \end{pmatrix}, \;
    \mathbf{D} = \begin{pmatrix}
    0 & 0 \\ 0 &  -\mathbf{K}_P
    \end{pmatrix}, 
\end{align*}
where $\mathbf{\hat{A}}~=~{\rm diag} (A_1,\ldots,A_N)$, $\mathbf{K}_P~=~k_P \eye$,
$\mathbf{K}_l = k_l \eye$
and $\eye$ is the unit matrix.
As Eq.~\eqref{eq:eom-lin-x} is a linear DDE with constant coefficients, it has eigenmodes of the form $\vec{x} = \vec x^\circ \frac{1}{2}(e^{\lambda t} + e^{\lambda^*t})$ \cite{amann2007some}.
The characteristic roots $\lambda$ and its complex conjugate can be determined by using the exponential ansatz $\vec x(t) = \vec x^\circ e^{\lambda t}$.
The eigenvalues can thus be determined from the characteristic equation 
\begin{align}
    \det\left( \lambda \mathbf{A} - (\mathbf{N} + \mathbf{D} e^{-\lambda \tau}) \right) = 0.
    \label{eq:characteristic}
\end{align}
As in conventional linear stability analysis, the stability is then encoded in the signs of the eigenvalues $\lambda$, i.e. all solutions of Eq.~\eqref{eq:characteristic}. 
The fixed point is linearly stable if the real part of all eigenvalues is negative and unstable if the real part of at least one eigenvalue is positive\cite{michiels2014ddestability}.
Solving the characteristic equation \eqref{eq:characteristic} is somewhat more challenging as in the non-delayed case. 
A reliable method to approximate the eigenvalue spectrum (i.e. the charcteristic roots of Eq.~\eqref{eq:characteristic}) is to use the Chebyshev collocation method\cite{breda2006chebyshev, jarlebring2008}.
It describes the state of the linearized delay differential equation $\vec{x}(\theta)$ in the time interval $[t- \tau, t]$ by discretzing at the so-called Chebyshev points $t_k = \cos{(\frac{k}{N_C}\pi)} \in [-1,1]$ with $k=0,\dots,N_C$ and $N_C$ giving the number of Chebyshev points and thus the resolution.
The DDE is now approximated as $y=[\vec{x}_0(t), \dots, \vec{x}_{N_C}(t)]^T$, where $\vec{x}_k(t) = \vec{x}(t - \frac{\tau}{2}(t_k + 1))$. 
One now has a $K(N_C+1)$-dimensional state vector $\vec{y}(t)$ instead of the $K$ dimensional original state $\vec{x}(t)$ transforming the linear delay differential equation to $\dot{\vec{y}}(t) = \mathbf{M_C} \vec{y}(t)$ with $\mathbf{M_C}$ given by
\newcommand*{\temp}{\multicolumn{1}{c|}{0}}
\[\arraycolsep=1.6pt\def\arraystretch{1.5}
\mathbf{M_C} =
\left(\begin{array}{c}
- \mathlarger{\frac{2 \; \mathbf{C_{M}}}{\tau} \otimes \mathbf{I_{K}}} \\[4pt]
\hline
\begin{array}{ccccc}
\mathbf{A^{-1}}\mathbf{D},  & \mathbf{0} & \dots & \mathbf{0} ,& \mathbf{A^{-1}}\mathbf{N}\\
\end{array}  \end{array}\right),
\]  
where $C_M$ is the Chebyshev differentiation matrix with the last row being deleted, $I_K$ the $K$-dimensional identity matrix and $\otimes$ the Kronecker product.
Note, the last row in $M_C$ is the original delay differential equation, while the others represent a spectral approximate of the time derviative at the Chebyshev nodes.

In addition to approximating the eigenvalue spectrum using Chebyshev discretization, one can evaluate the points in parameters space, where an eigenvalue might pass the imaginary axis.
These points are candidates for switches in stability, since it requires that at least one eigenvalue passes the imaginary axis for the stability of the fixed point to change.
This can be done by casting the characteristic equation given in Eq.~\eqref{eq:characteristic} as the eigenvalue problem 
\begin{align}
    \left( \lambda\mathbf{A} - (\mathbf{N} + \mathbf{D} e^{ - \lambda \tau}) \right) \vec{\nu}(\lambda) = 0
\end{align}
which is a transcendental equation with infinitely many solutions and is thus in general more challenging as in the case for ordinary differential equations.
Leveraging the structure of the system, we can write the equation in higher orders of $\lambda$ by introducing $\vec{\nu}(\lambda) = [\vec{u}, \lambda\vec{u}]^T$ with $\lambda\vec{u}$ giving the frequency deviation in Laplace domain. This leads to 
\begin{align}
    \left(\lambda^2 \mathbf{\hat{A}} + \lambda \mathbf{K}_l + \mathbf{L} \right) e^{\lambda \tau} \vec{u}(\lambda) = -\lambda \mathbf{K}_P \vec{u}(\lambda).
\end{align}
Solving for the eigenvalues $\sigma(\lambda)$ of the matrix $\mathbf{M}(\lambda) = \left(\lambda^2 \mathbf{\hat{A}} + \lambda \mathbf{K}_l + \mathbf{L} \right)$ we can write
\begin{align}
    \sigma(\lambda) e^{\lambda\tau} = - \lambda k_P. 
\end{align}
Since we want to determine the point in parameter space where $\lambda$ has a vanishing real part, we substitute $\lambda = i \eta$ and obtain $k_P$ as 
\begin{align}
    k_P = - \frac{\sigma(i \eta) e^{i \eta \tau}}{i \eta}.\label{eq:crit_kp}
\end{align}
Since we know that the $k_P$ of interest are not complex, we have to determine the $\eta_c$ for which the imaginary part of the right-hand side of Eq.~\eqref{eq:crit_kp} vanishes by scanning over a grid of $\eta$ and solving the eigenvalue problem.
To accurately record where the imaginary part vanishes, the individual values for different $\eta$ have to be sorted to obtain the curves that describe how the right-hand side of Eq.~\eqref{eq:crit_kp} change as a function of $\eta$. 
Since the eigenvalues and associated eigenvectors change only slightly for close $\eta$ values, we have used them to sort the different eigenvalues $\sigma$ for a chosen $\eta$.
A more detailed description of this method and closely related method can be found in refs.~\cite{bottcher2020time, otto2014extension} and ref.~\cite{ramirez2019approach}, respectively.

\begin{figure}[tb]
    \centering
    \includegraphics[width=\columnwidth]{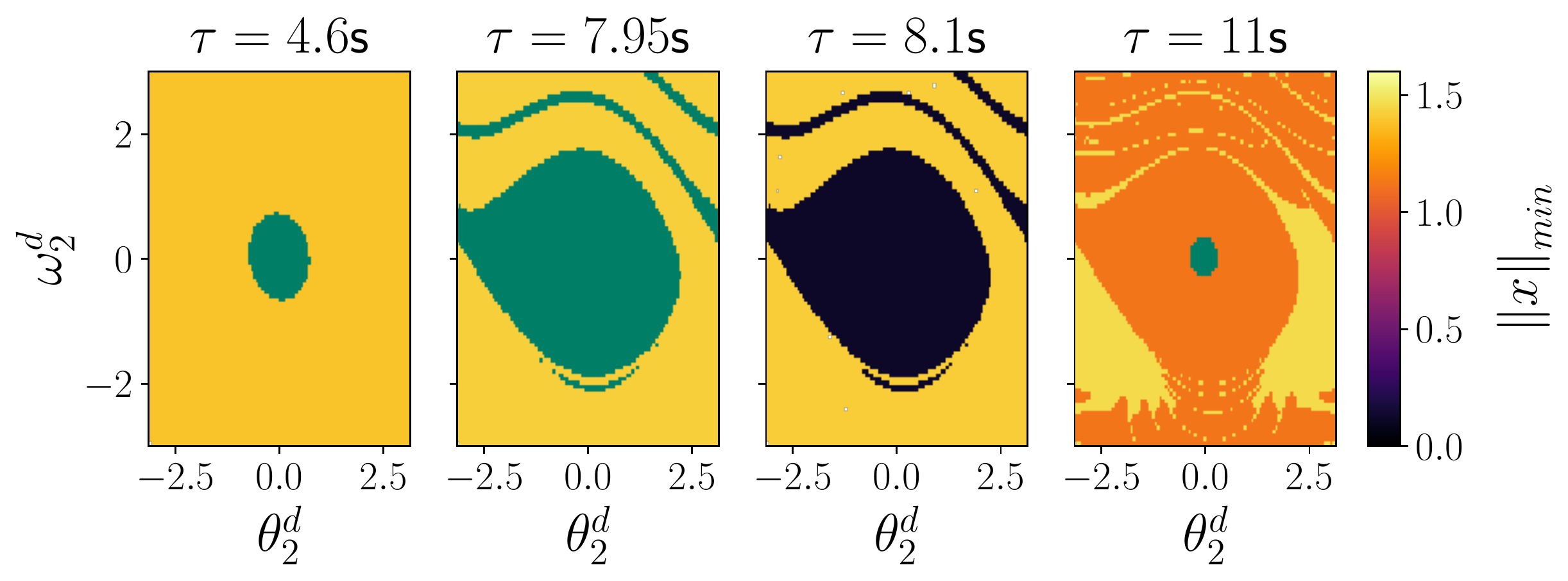}
    \caption{Basin of the fixed point and surrounding limit cycles of the delayed differential equation describing power grid dynamics \eqref{eq:full_dde} for different values of the delay $\tau$. 
    Simulations were performed by disturbing the stationary frequency deviation and power phase angles of area two by $\omega_2^d$ and $\theta_2^d$, resprectively.
    From left to right, the values for $\tau$ were chosen as $4.6$s, $7.95$s, $8.1$s and $11$s.
    The basin of the fixed point corresponding to operation of the reference system is shown in green. 
    The color scale describes the minimal 2-norm of the distance of the limit cycle dynamics to the fixed point $\|x\|_{\rm min}$ showing the coexistence of different limit cycles.}
    \label{fig:delay_basins}
\end{figure}

\subsection{Global stability}

The global stability of the power system model given in Eq.~\eqref{eq:2ndorder} is quantified in terms of the basin of attraction of the desired fixed point. 
However, measuring the size of the basin becomes challenging for delayed differential equations. 
It is not sufficient to choose a point in phase space to specify the initial state of the system.
Instead, the function $\vec x_0(t')$ must be specified for $t' \in [t_0-\tau,t_0]$.
While it might be useful to taylor the specific initial funciton to the application at hand or sampling from a reasonable set of initial functions\cite{leng2016basin}, the choice of initial functions is ultimately arbitrary, which makes it hard to make a general statement on the global stability of a attractor.
To get around this problem, an efficient method to assess global stability in delayed systems has been proposed in \cite{scholl2019time,scholl2020norm}. 

In principle, one chooses a suitable initial function segment and solves the delay differential equation.
We will choose a constant past given by setting the state vector $x_{0}(t')$ to a constant value for $t'\in [t_0 - \tau, t_0]$.
Keep in mind that a larger delay with the same initial constant effectively represents a larger disturbance. 
Thus, it is useful to evaluate the 2-norm, which can be seen as the energy of a specific disturbance, of the initial function segment instead of the randomly chosen value that gives the constant past. 
To quantify how stable a fixed point is, the initial function with the smallest norm that does not result in the dynamics relaxing to the considered fixed points is of interest. 
This value gives the primary attractor radius which is still a bad approximate for the size of the basin of attraction since it only considers constant function segments. 
Subsequently, all simulations that did not return to the considered fixed points are used to get a better estimate for the size of the basin. 
By cutting them up into all possible segments of length $\tau$ and measuring the 2-norm of the segments and keeping the minimum,  the secondary attractor radius can be calculated. 
Choosing the minimum of the primary and secondary attractor radius, gives a measure for the smallest possible disturbance that results in the dynamics not relaxing to the considered fixed point.

While measuring the basin has its limitation\cite{schultz2017potentials} and this is even more true for systems that include delayed dynamics\cite{leng2016basin}, knowledge of the attractor radius $R_a$ can be used to examine how the approximate size of the basin changes for different parameters or more specifically which smallest disturbance leads to the attractor not being recovered after a disturbance.

\subsection{Results}

\begin{figure}[tb]
    \centering
    \includegraphics[width=\columnwidth]{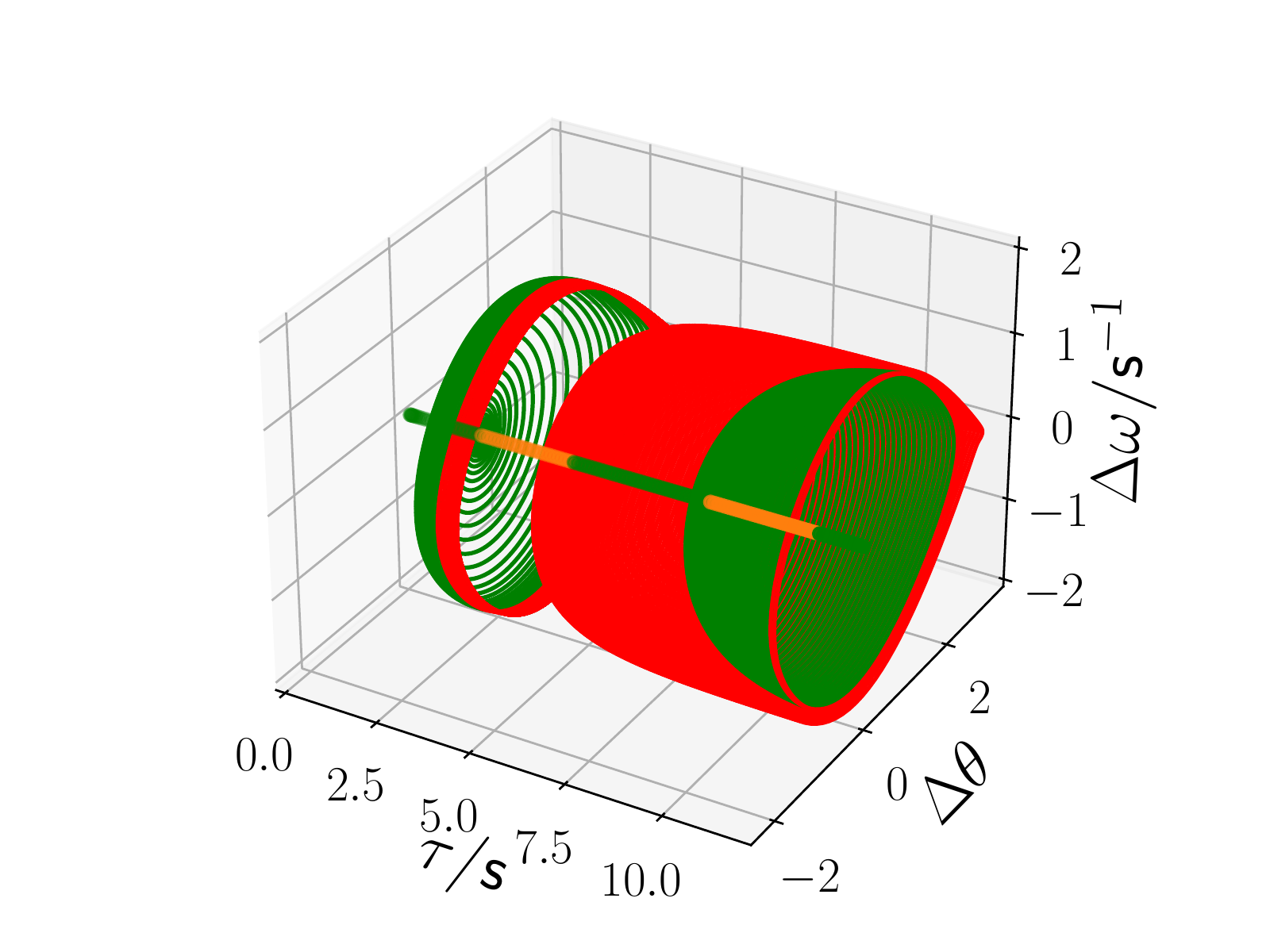}
    \caption{Bifurcation diagram of the delayed dynamical system \eqref{eq:2ndorder}. The colorcode indicates the linear stability of the respective attractor with green encoding linear stability.
    Red and orange show that there are one and two unstable eigenvalues, respectively, such that the dynamics is linearly unstable.
    We observe a set of consecutive sub- and super-critical Hopf-bifurcations and the coexistience of different limit cycles.
    }
    \label{fig:3dbiff}
\end{figure}

We now compute the local and global stability in a power grid model consisting of two areas with homogeneous parameters.

This model system captures essential physical processes, in particular the interplay of inter-area oscillations and control systems, but still allows for a comprehensive visual analysis. 
Choosing $k_P~=~0.0625$, $k_l~=~0.025$, $A~=~1$s, $C_{1,2}=0.5$ and $P_{0, \{1,2\}}~=~\pm0.0625$, we study the resulting dynamics and stability as a function of the delay time $\tau$. 
We initialize the simulation by setting the initial function to the constant state $\vec{x}_0 = [\omega_1^0, \omega_{2}^0, \theta_1^0, \theta_2^0] = [0, \omega_2^d, \theta_1^*, \theta_2^* + \theta_2^d]$ with $\omega_2^d$ and $\theta_2^d$ being the difference to the stationary values of the frequency deviation and the power phase angle of the second area, respectively.

Figure~\ref{fig:delay_basins} provides a first visual overview of the global stability for four different values of $\tau$.
In particular, Fig.~\ref{fig:delay_basins} shows the constant values chosen to define the function segments that serve as the initial state of the DDE given in Eq.~\eqref{eq:full_dde}.

The fixed point is linearly stable for $\tau=4.6$s, but has a rather small basin of attraction confined by an unstable limit cycle. Increasing the delay to $\tau=7.95$s, both the limit cycle and the basin grow. 
A further small increase of the delay to $\tau=8.1$s makes the desired fixed point unstable. 
For many initial function segments, the system now relaxes to a stable limit cycle with small amplitude indicating that a supercritical Hopf bifurcation took place. 
Finally, the fixed point regains its stability for $\tau=11$s, again with a comparatively small basin of attraction. 
Outside of this basin, the system relaxes to two different limit cycles for the chosen set of values defining the constant initial function segments. 
This highlights a similar shell structure of different attractors around the trivial fixed point as also observed in the continuously kicked system.

\begin{figure}[tb]
    \centering
    \includegraphics[width=\columnwidth]{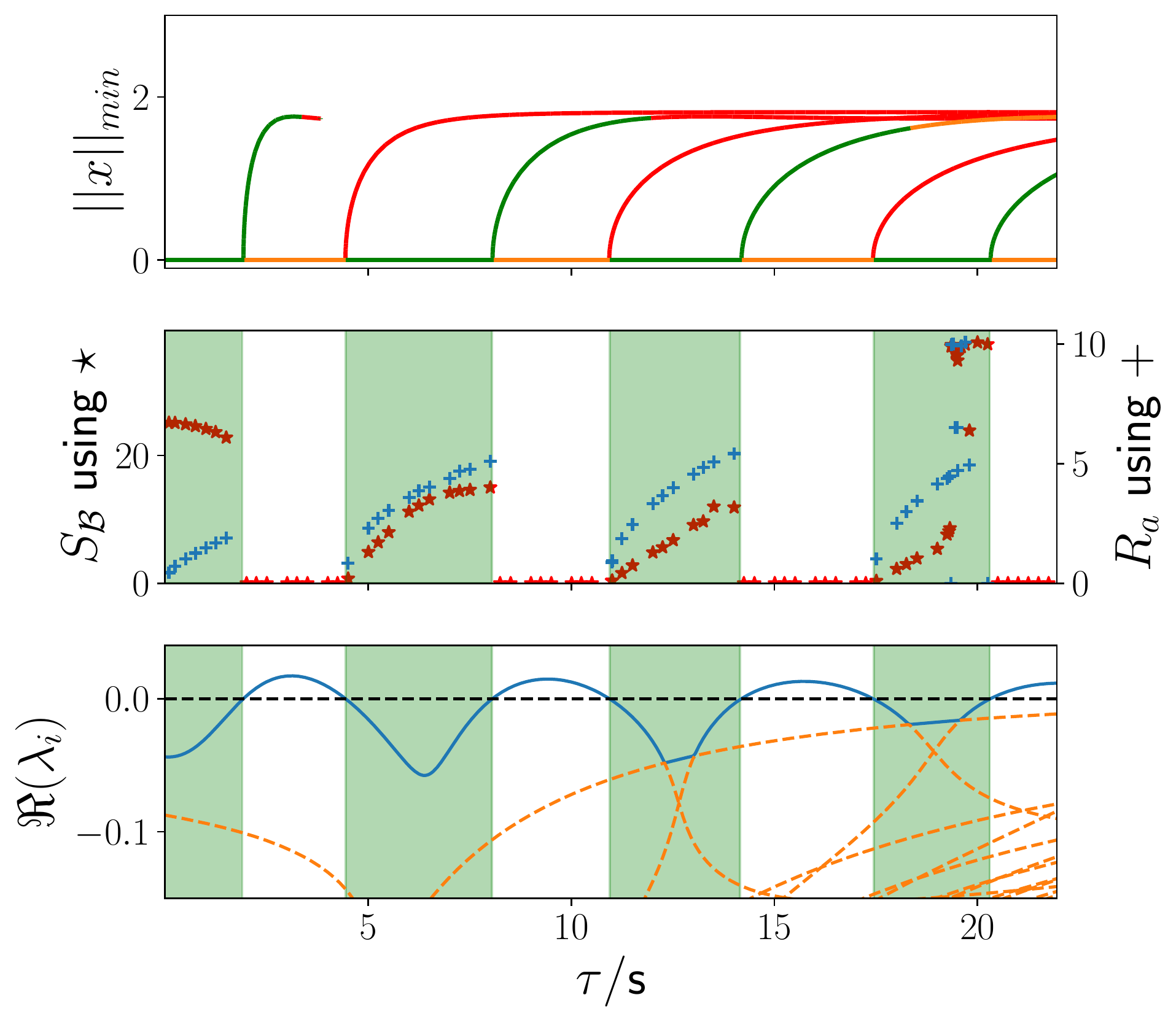}
    \caption{Global and local stability of the delayed dynamical system \eqref{eq:2ndorder} as a function of the delay $\tau$. Top: Minimal 2-norm $\|x\|_{\rm min}$ of the emerging limit cycles. Color indicates the number of unstable eigenvalues with green, orange and red indicating zero, one and two unstable roots, respectively. 
    Middle: Size of the basin of attraction of the fixed point given by the volume in phase space $S_\mathcal{B}$ (red stars) and the attractor radius $R_A$ (blue +). 
    Bottom: Dominant eigenvalue of the Jacobian evaluated at the fixed point in blue. The orange dashed lines shows the next  eigenvalues. Green shaded areas show $\tau$-regions where the fixed point is linearly stable.}
    \label{fig:biff_eigvals_tauscan}
\end{figure}

To gain further insight into the interaction of limit cycles and fixed points, we perform a bifurcation analysis using the software DDE-BIFTOOL \cite{ddebiftool} to complement the linear stability analysis of the fixed points described in Sec.~\ref{sec:power_linstab}.
The results in Fig.~\ref{fig:3dbiff} shows a series of consecutive sub- and super-critical Hopf-bifurcations, similar to the examples presented in Sec.~\ref{sec:kicked_system}. Without delay, $\tau=0$s, the fixed point is linearly stable. As $\tau$ increases, stability is lost in a supercritical Hopf bifurcation and a stable limit cycle emerges. After a further increase of $\tau$, stability is regained in a sub-critical Hopf bifurcation where an unstable limit cycle emerges. This scenario repeats itself when the delay $\tau$ is increased further, leading to a pronounced shell structure in phase space.

We can now provide a comprehensive analysis of the local and global stability of the desired fixed point as a function of the delay $\tau$. Figure~\ref{fig:biff_eigvals_tauscan} compares the location of limit cycles, the size of the basin of attraction and the linear stability of the fixed point.
As the delay $\tau$ increases, the fixed point repeatedly switches from stable to unstable and back. 
Mathematically, this corresponds to a set of super- and subcritical Hopf bifurcations as explained above. 
Physically, the instability can be explained as a resonance effect \cite{schafer2015decentral}. 
The delayed control amplifies the inter-area eigenmode instead of damping it. 
Hence, regions of instability are found where the delay $\tau$ matches an integer multiple the period of the eigenmodes.

The limit cycles generally increase in size with the delay $\tau$. This leads to an opposite behavior of linear and global stability. Within the regions of stability, the basin typically grows monotonically with $\tau$. Hence, basin stability is largest at the bifurcation, when the fixed point becomes unstable again. 
Remarkably, we find the highest value of the basin size for $\tau \approx 20$s, at the edge of the fourth stability region.

\begin{figure}[tb]
    \centering
    \includegraphics[width=\columnwidth]{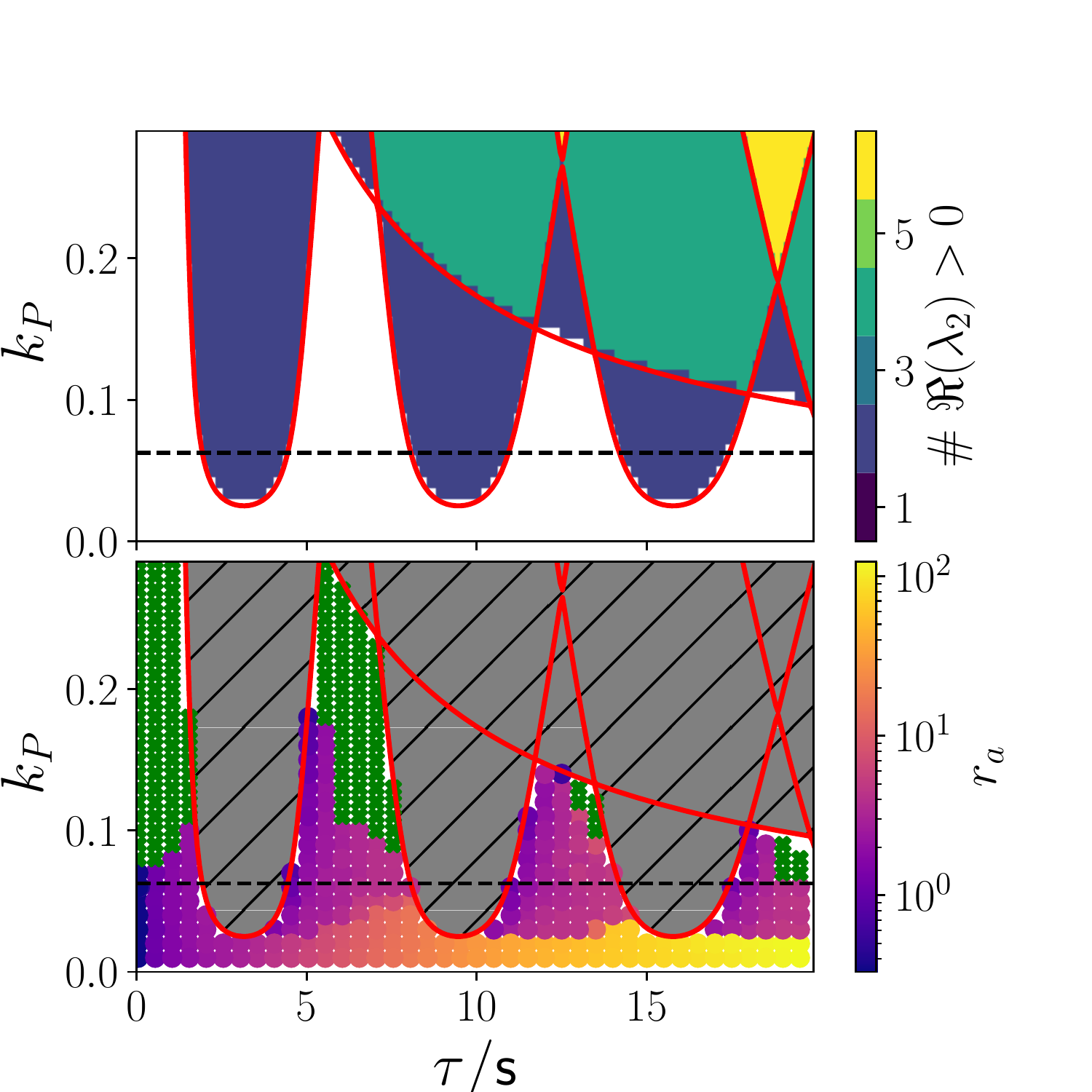}
    \caption{Stability chart as function of $k_P$ and $\tau$. The results from the linear stability analysis via Chebyshev Discritization and attractor radius $r_a$ are shown in the top and bottom panel, respectively.
    Parameter combinations for which all simulations with the chosen initial functions returned to the fixed point are shown as green crosses in the bottom panel.
    Red lines show critical curves in parameter space on which an eigenvalue has a zero real part that were determined by the method described above.
    The $k_P$ value used in Fig.~\ref{fig:3dbiff} and Fig.~\ref{fig:biff_eigvals_tauscan} is indicated by the black dashed lines. 
    Regions in parameter space where no stable fixed point is found are indicated by the gray hashed areas.
    }
    \label{fig:cheby_attr}
\end{figure}

Since the value of $k_P$ is a choice that is given by the control design, it is important to understand how it affects the stability of the desired fixed point.
The stability chart in Fig.~\ref{fig:cheby_attr} shows both the results of the linear stability analysis and the results obtained by finding the attractor radius $r_a$. 
The results of the linear stability analysis is given both by evaluating the approximate eigenvalue spectrum using Chebyshev discretization and finding critical curves with at least one eigenvalue having a vanishing real part, while the attractor radius was evaluated as described.
The changes in stability are always accompanied by a change (i.e., increase or decrease) of the unstable eigenvalues which is characteristic for a Hopf-bifurcations. 
The dashed black lines in Fig.~\ref{fig:cheby_attr} show the parameter used in Fig.~\ref{fig:biff_eigvals_tauscan} and the gray hatched area in Fig.~\ref{fig:cheby_attr} gives the region in parameter space where the fixed point is not stable. 
Note, that for larger $k_P$ an additional critical curve that limits the maximal delay that the system can tolerate. 
This mode, in addition to the repeating critical curves that give rise to the changing sub- and super-critical Hopf-bifurcations, shapes the stability chart and determines regions with stable and unstable fixed points. 
The attractor radius (see bottom panel in Fig.~\ref{fig:cheby_attr}) shows a similar behavior as we have seen previously. 
It increases between two regions where no stable fixed point can be found with increasing delay $\tau$ and reaches a maximum slightly before stability is lost.
Additionally, there are regions in parameter space were no other stable attractor than the stable fixed point exists (see green crosses in Fig.~\ref{fig:cheby_attr}). 
This can be attributed to a bifurcation of the limit cycles, which highlights again that the stability of a fixed point depends crucially on how different attractors interact to shape the stability chart.

\FloatBarrier

\section{Discussion and Conclusion}

Summarizing, we have demonstrated how local and global stability may give contrarian results: Large basin of attraction volumes may coincide with vanishing linear stability at a Hopf bifurcation. Critically, we have shown that this effect is not limited to basic toy systems but also emerges in more complex anharmonic oscillators and in (delayed)  power grid dynamics. 

With our work we have substantially expanded upon earlier advances, which noted the "perfect stability" in delayed power systems \cite{schafer2015decentral,schafer2016taming}. We have stressed the critical role of Hopf bifurcations and the interplay between growing unstable limit cycle orbits and the basin of attraction of a stable fixed point. 
To further solve the apparent paradox of two conflicting stability statements, we might hypothesize an analogy to phase transitions, also observed in bifurcations, the "critical slowing down" \cite{tredicce2004critical}. When the dynamical system approaches the bifurcation, the unstable limit cycle has its largest share of the phase space and all initial conditions within the cycle have to converge to the stable fixed point. However, this convergence will be slower than for smaller basins (as indicated by vanishing linear stability). Hence, we observe a transition from an initial steep (high linear stability) but narrow (small global stability) basin to a flat (low linear stability) and wide (high global stability) basin. 

Our results are very interesting both from a dynamical system perspective and from an application point of view: If system parameters are well controlled and eventual convergence is the main goal, operating a dynamical system close to the unstable bifurcation point could be desirable, as many perturbations, even large ones, will still converge eventually to the stable state. Meanwhile, if a quick convergence is desired, e.g. in power grid control, operation should be far away from the bifurcation point, while keeping in mind that large deviations in phase space are very dangerous and could drive the system away from its desired state, hence, a tight control of the system to ensure its proximity to the fixed point is necessary at all times. 

With this contribution, we have shed some light on conflicting statements from linear and basin stability. Still, many open questions remain. In the future, it would be interesting to observe whether this contrarian effect of local and global stability can also be shown for other bifurcations and basins of limit cycles (in addition to basins of fixed points). 
Additionally, future studies could include an extension of the power grid model by adding more details of the currently used control mechanisms and more complex topologies to see how the local and global stability behave. 
Here, larger systems with heterogeneously distributed inertia are of special interest, since future power system need to deal with renewable generation by solar and wind being placed at locations with the highest potential yield.
Thus, regions with a high share of fluctuating renewables supply a considerably low amount of inertia. 
Knowing how the local and global stability is affected by these developments and how to choose control parameters or delays that guarantee a stable system, could play a vital role in designing a robust future power system.

\begin{acknowledgments}
We thank Leonardo Rydin Gorj\~ao for stimulating discussions and proofreading the manuscript.
We gratefully acknowledge support from the German Federal Ministry of Education and Research (BMBF) via the project CoNDyNet2 (grant no. 03EK3055B), from the Helmholtz Association via the grant no.~VH-NG-1727 and the project \textit{Uncertainty Quantification -- From Data to Reliable Knowledge (UQ)} with the grant no.~ZT-I-0029, and the Deutsche Forschungsgemeinschaft (DFG, German Research Foundation) via grant no.~ 491111487.
P.C.B. acknowledges the support of the German Federal Ministry for Economic Affairs and Energy (BMWi) via the project DYNAMOS (grant no. 03ET4027A) in which parts of the methodology for the stability analysis of the power grid model were developed at the DLR Institute of Networked Energy Systems.

Parts of the simulations were performed at the HPC Cluster CARL, located at the University of Oldenburg (Germany) and funded by the DFG through its Major Research Instrumentation Programme (INST 184/157-1 FUGG) and the Ministry of Science and Culture (MWK) of the Lower Saxony State.
\end{acknowledgments}

\input{main.bbl}

\end{document}

%% file: main.bbl
%